\documentstyle[12pt,a4]{article}

\begin{document}
\thispagestyle{empty}
\vspace*{-1.5cm}
\hfill {\small KL--TH 96/11} \\[8mm]

\begin{center}
{\Large\bf M-branes, anti-M-branes\\
and nonextremal black holes}
\\[14mm]
\renewcommand{\thefootnote}{*}
{\large Jian--Ge Zhou\raisebox{0.8ex}{\small a,}\footnote {Email:
jgzhou @ physik.uni-kl.de},
H.~J.~W. M\"uller--Kirsten\raisebox{0.8ex}{\small a},\\
J.--Q. Liang\raisebox{0.8ex}{\small a,b}, 
F. Zimmerschied\raisebox{0.8ex}{\small a,} }
\\
\raisebox{0.8ex}{\small a}  Department of Physics, \\
University of Kaiserslautern, D--67653 Kaiserslautern,
Germany \\
\raisebox{0.8ex}{\small b}  Institute of Theoretical Physics, \\
Taiyuan, 030006 Shanxi, P.\ R.\ China \\
Institute of Physics, Academia Sinica,\\
Beijing 100080, P.\ R.\ China \\
\end{center}

\section*{Abstract}
An M-brane and anti-M-brane scheme is proposed to 
study nonextremal $4D$ and $5D$ black holes. The improved nonextremal
intersecting M-brane solutions proposed here, involve two sets of harmonic functions. 
The constraints among the pressures are found, and new features 
in the M-brane and anti-M-brane picture are demonstrated, which resolve the 
discrepancy in the number of free parameters in the D-brane picture.
In terms of the ``numbers'' of M-branes and anti-M-branes, the prefactors of 
the entropies are found to be model independent, and the Bekenstein-Hawking
entropy assumes the duality invariant form which is consistent with the
microscopic explanation of the black hole entropy.

\vspace{2 cm}
\noindent
PACS: 04.70 Bw, 11.25.-w, 04.20.Jb \\
Keywords: black holes, p-branes, M-theory, string theory
\newpage

\section{Introduction}
Recently, more and more evidence has accumulated to show that the best candidate for a
unified theory underlying all physical phenomena is no longer $D = 10$ 
superstring theory but rather $D = 11$ M-theory which generalizes known
string theories \cite{1}, and $D = 11$ supergravity can be regarded as
a low-energy effective field theory of the fundamental M-theory. The precise
formulation of M-theory is not clear, but 2-branes and 5-branes play an
important role in the $D = 11$ supergravity theory owing to the 4-form
field strength $F_4$ \cite{2}. The supersymmetric BPS-saturated p-brane
solutions of low-dimensional theories can be understood as reductions of the
basic $D = 11$ M-branes, i.e. as 2-branes \cite{3} and 5-branes \cite{4} and their
combinations \cite{5}. The generalization to a number of different harmonic
functions specifying intersecting BPS-saturated M-branes and a construction
of new intersecting p-brane solutions in $D \le 11$ were given in \cite{6}.
It has been shown that there exists a simple harmonic function rule which 
governs the construction of composite supersymmetric p-brane solutions in
both $D = 10$ and $D = 11$, and a separate harmonic function is assigned 
to each constituent $\frac12$ supersymmetric p-brane \cite{6}. Later, the
extremal supersymmetric BPS-saturated intersecting M-brane solution was 
generalized to the nonextremal case \cite{7}. The resulting nonextremal
intersecting M-brane solutions are no longer supersymmetric, but they can be 
constructed as a deformation of extremal solutions, parametrised in terms of
several one-center harmonic functions and the Schwarzschild solution,
parametrised in terms of $m$. The nonextremal configurations of intersecting M-branes 
can be related to nonextremal black holes in dimensions $D = 4$, $D = 5$ and 
$6 \le D \le 9$, by dimensional reduction along internal M-brane directions.
In \cite{7}, the nonextremal black holes were explained as nonextremal 
intersecting M-branes which has nothing to do with the D-brane-antibrane 
picture \cite{8,9,10}. On the other hand, the statistical origin of the 
Bekenstein-Hawking entropy of certain extremal black holes in string theory
can be elucidated by the D-brane technique \cite{11}-\cite{18}, and in particular, the
nonextremal black holes can be decomposed into a collection of D-branes and
anti-D-branes \cite{8,9,10}. Then one may ask whether it is possible to 
discuss the nonextremal black holes in M-theory from the M-brane and 
anti-M-brane approach by trading the parameters of the general solution for
the ``numbers'' of M-branes and anti-M-branes. As shown below, this can be done
only by relating the unconstrained pressures, ADM mass and gauge charges of 
the black hole to the ``numbers'' of a collection of noninteracting
constituent branes and antibranes. In the D-brane picture, the number of free
parameters should consist of the ``numbers'' of D-branes and anti-D-branes
plus the values of moduli \cite{19}, but it turns out that for the black
hole solutions in string theories, when the ``numbers'' of D-branes and 
anti-D-branes are given, the values of moduli will be fixed, that is, they 
are determined by the actual number of branes of each type \cite{8}. As a 
result, a dicrepancy in the number of free parameters appears \cite{8}. 
In fact, seen from the black hole picture, the values of moduli are neither 
completely arbitrary, nor fixed, since if they are too large, the solutions
become classically unstable \cite{20}. However, how the above
contradiction can be reconciled in the framework of D-branes is unclear.

In the present paper, the D-brane and anti-D-brane picture is extended to
M-theory, i.e., M-branes and anti-M-branes. The nonextremal $4D$ and $5D$
black holes obtained upon toroidal compactification from the nonextremal
intersecting M-brane solutions are first identified as the composition of
M-branes and anti-M-branes. In a unified frame of $D = 11$ M-theory, the
``numbers'' of M-branes and anti-M-branes are defined by matching thermodynamic
properties of the black hole to thermodynamic properties of a collection
of noninteracting M-branes and anti-M-branes. In our improved nonextremal
intersecting M-brane solutions, two sets of harmonic functions $\sigma_i
(\omega_i)$ and $\tilde{\sigma}_i(\tilde{\omega}_i)$ are first introduced,
and the field thrength $F_4$ for 2-branes is different from that in \cite{7}
where only one set of harmonic functions $T^{-1}_i$ was given and the other set of
complicated function $T_i^{'-1}$ were not harmonic ones. From our construction 
for the field strength $F_4$, it is easy to calculate explicitly the electric
charges, in fact the electric and magnetic charges can be evaluated in a systematic 
way in eleven dimensions. It is first found that there exist constraints 
among the pressures, and only the unconstrained pressures, ADM mass and
gauge charges match the ``numbers'' of M-branes and anti-M-branes. Unlike the
D-brane and anti-D-brane picture where the moduli are only functions of
the ``numbers'' of D-branes and anti-D-branes \cite{8}-\cite{10}, there are new 
features in the M-brane and anti-M-brane picture: 1) as shown in section
2.1, there are constraints among the moduli, and some moduli can be chosen as
free parameters; 2) the moduli depend not only on the ``numbers'' of M-branes and 
anti-M-branes, but also on the nonextremality parameter $m$. The reason for
this is that the masses of M-branes are not independent parameters, instead they
are proportional to the common mass parameter $m$, and the configurations of the
nonextremal intersecting M-brane solutions should be viewed as ``bound-state''
configurations. The constraints among the pressures and moduli found here are different
from those in \cite{10}, where the appearance of their constraints are due to
introducing the auxiliary SU(8) multiplet via E(7) symmetry. In the present M-brane
picture, when the ``numbers'' of M-branes and anti-M-branes are kept, the
values of moduli can not be fixed completely, which resolves the discrepancy
in the number of free parameters in the D-brane picture. As we know, when the black
hole entropies are expressed in terms the original parameters of the solutions,
the prefactors of the entropies depend on the dimensions of the black hole \cite
{7,21}. In terms of the ``numbers'' of M-branes and anti-M-branes, the Bekenstein-Hawking entropy takes the duality invariant form which is consistent with the 
microscopic explanation of the black hole entropy, and the prefactors of
the entropies are found to be $2\pi$, which is independent of the dimensions
of the black holes. In \cite{8,9}, only one model was discussed in each paper, 
and the authors fit their conventions in different ways so that the prefactors are
$2\pi$. However, in \cite{21}, since no unified frame was exploited, the
prefactors are $2\pi, 4\pi$ respectively in $D = 4,5$ dimensions. Here we
study different models in a {\it unified} frame of $D = 11$ M-theory; thus the 
universality of the prefactors is nontrivial. The same prefactors of the 
entropies between two black holes in the same dimensions obtained from two
different nonextremal intersecting M-brane configurations imply that these
two resulting black hole backgrounds are related by the symmetry transformation
of the M-theory, which is a combination of T-duality and SL(2,Z) symmetry of 
the $D = 10$ type IIB theory lifted to $D = 11$ M-theory. Even though the
general formula for the statistical entropy cannot be derived from a counting
of states, this can be done in certain limits corresponding to near-extremal
black holes. By exploiting U-duality, one can interchange the different
branes, then the resulting equation for microscopic entropy is duality
invariant, which agrees with different nonextremal limits.

The layout of the paper is as follows. In the next section we consider two
$4D$ nonextremal black holes obtained upon toroidal compactification from
the nonextremal intersecting M-brane configurations $2\perp2\perp5\perp5$
and ``boosted'' $5\perp5\perp5$, and define the ``numbers'' of M-branes and
anti-M-branes in a unified frame of $D = 11$ M-theory. We display the
constraints among the pressures and moduli, and express the Bekenstein-Hawking 
entropies in terms of the ``numbers'' of M-branes and anti-M-branes. 
In section 3, we discuss two nonextremal $5D$ black holes obtained from the
configurations $2\perp2\perp2$ and ``boosted'' $2\perp5$ in a way similar to
that of section 2. In section 4, we explain the microscopic origin of the 
Bekenstein-Hawking entropy for the nonextremal black holes. Finally, in
section 5, we present our conclusions.

\section{$D = 4$ nonextremal black holes}
We consider nonextremal $4D$ black holes which can be obtained upon toroidal
compactification from two kinds of nonextremal intersecting M-brane
configurations, i.e. $2\perp2\perp5\perp5$ and the ``boosted'' 
$5\perp5\perp5$ \cite{7,22}.

\subsection{The $2\perp2\perp5\perp5$, $D = 11$, solution}
From the algorithm which leads to the nonextremal version of a given extremal
intersecting M-brane solution \cite{7}, the $2\perp2\perp5\perp5$
configuration in $D = 11$ M-theory is described by

\begin{eqnarray}
ds^2_{11} = (\sigma_1\sigma_2)^{\frac13} (\omega_1\omega_2)^{\frac23}
\Big[ -(\sigma_1\sigma_2\omega_1\omega_2)^{-1} 
\left(1-\frac{2m}{R}\right) dt^2 \nonumber \\
+ (\sigma_1\omega_1)^{-1} dz^2_1 + (\sigma_1\omega_2)^{-1} dz^2_2 + 
(\sigma_2\omega_1)^{-1} dz^2_3 + (\sigma_2\omega_2)^{-1} dz^2_4 \nonumber\\
+ (\omega_1\omega_2)^{-1} (dz^2_5 + dz^2_6 + dz^2_7) + 
\left(1-\frac{2m}{R}\right)^{-1} dR^2 + R^2d\Omega^2_2 \Big] 
\label{1}
\end{eqnarray}
The associated gauge field which must be a 4-form is given by
\[
F_4 = F_4(\sigma_1) + F_4(\sigma_2) + F_4(\omega_1) + F_4(\omega_2)
\]
with
\begin{eqnarray}
F_4(\sigma_1) & = & 3 dt \wedge (\sigma_1^{-2} d\tilde{\sigma}_1) \wedge
dz_1 \wedge dz_2 \nonumber \\
F_4(\sigma_2) & = & 3 dt \wedge (\sigma_2^{-2} d\tilde{\sigma}_2) \wedge
dz_3 \wedge dz_4 \nonumber \\
F_4(\omega_1) & = & 3(*d\tilde{\omega}_1) \wedge dz_2 \wedge dz_4 
\nonumber \\
F_4(\omega_2) &=& 3(*d\tilde{\omega}_2) \wedge dz_1 \wedge dz_3
\label{2}
\end{eqnarray}
where the dual form * is defined in the asymptotically flat 3-dimensional 
transverse space, and
$\sigma_i(\tilde{\sigma}_i), \omega_i(\tilde{\omega}_i)$ are harmonic
functions corresponding to the 2-branes, 5-branes respectively,
\begin{eqnarray}
\sigma_i &=& 1 + \frac{2m \sinh^2 \alpha_i}{R} , \quad \tilde{\sigma}_i = 1 + \frac{m \sinh 2 \alpha_i}{R} \nonumber \\
\omega_i &=& 1 + \frac{2m \sinh^2 \beta_i}{R} , \quad \tilde{\omega}_i = 1 + \frac{m \sinh 2 \beta_i}{R}, \quad i = 1,2
\label{3}
\end{eqnarray}
where $\alpha_i$ and $\beta_i$ are related to the electric and magnetic 
charges and $m$ is the nonextremality parameter. Here we note that our improved
nonextremal intersecting M-brane solution for the field strength $F_4(\sigma_i)$
is different from that in \cite{7}.

To obtain the electric charges, we calculate the quantities $^{\hat{*}}
F_4(\sigma_1), ^{\hat{*}}F_4(\sigma_2)$, where $\hat{*}$ denotes the $D = 11$
Hodge dual. After a series of calculational steps, one has
\begin{eqnarray}
^{\hat{*}}F_4(\sigma_1) = m (\sinh 2 \alpha_1) \epsilon_2 \wedge dz_3
\wedge dz_4 \wedge dz_5 \wedge dz_6 \wedge dz_7 \nonumber \\
^{\hat{*}}F_4(\sigma_2) = m (\sinh 2 \alpha_2) \epsilon_2 \wedge dz_1
\wedge dz_2 \wedge dz_5 \wedge dz_6 \wedge dz_7 
\label{4}
\end{eqnarray}
where $\epsilon_2$ is the area 2-form of $S^2$. Then the electric and magnetic 
charges can be defined as
\begin{eqnarray}
q_e^{(1)} &=& \frac{1}{\sqrt{2}\kappa} \int \,^{\hat{*}}F_4(\sigma_1) = \frac{4\pi m \sinh2\alpha_1}
{\sqrt{2}\kappa} l_3l_4V_3 \nonumber \\
q_e^{(2)} &=& \frac{1}{\sqrt{2}\kappa} \int \,^{\hat{*}}F_4(\sigma_2) = 
\frac{4\pi m \sinh2\alpha_2}{\sqrt{2}\kappa} l_1l_2V_3 \nonumber \\
q_m^{(1)} &=& \frac{1}{\sqrt{2}\kappa} \int F_4(\omega_1) = \frac{4\pi m \sinh2\beta_1}
{\sqrt{2}\kappa} l_2l_4 \nonumber \\
q_m^{(2)} &=& \frac{1}{\sqrt{2}\kappa} \int F_4(\omega_2) = \frac{4\pi m \sinh2\beta_2}
{\sqrt{2}\kappa} l_1l_3
\label{5}
\end{eqnarray}
where $\kappa^2/8\pi$ is Newton's constant in 11 dimensions, and in the
above calculation, we have assumed the internal coordinates $z_i, i = 1,...,4$, are 
periodically identified with period $l_i$, and $z_i, i = 5,6,7,$ are
identified with period $(V_3)^{\frac13}$. As a result, the solution (\ref{1})
has ten parameters: $m, \alpha_i, \beta_i, i=1,2, l_j, j=1,...,4$ and $V_3$.

The ADM mass of the solution (\ref{1}) is \cite{7,23}
\begin{equation}
M_{ADM} = \frac{2\pi m}{\kappa^2} l_1 l_2 l_3 l_4 V_3 (\cosh 2\alpha_1 +
\cosh 2 \alpha_2 + \cosh 2 \beta_1 + \cosh 2 \beta_2)
\label{6}
\end{equation}

In eleven dimensions, besides the ADM mass and the gauge charges, the black 
hole is also characterized by pressures which are related to the asymptotic
fall-off of the coefficents of $(dz_i)^2, i=1,...,5$, in the solution 
(\ref{1}). These pressures $P_i$ are given by \cite{8}
\begin{eqnarray}
P_1 &=& \frac{2\pi m}{\kappa^2} l_1 l_2 l_3 l_4 V_3 (-2\cosh 2\alpha_1 +
\cosh 2 \alpha_2 - \cosh 2 \beta_1 + 2\cosh 2 \beta_2) \nonumber \\
P_2 &=& \frac{2\pi m}{\kappa^2} l_1 l_2 l_3 l_4 V_3 (-2\cosh 2\alpha_1 +
\cosh 2 \alpha_2 + 2\cosh 2 \beta_1 - \cosh 2 \beta_2) \nonumber \\
P_3 &=& \frac{2\pi m}{\kappa^2} l_1 l_2 l_3 l_4 V_3 (\cosh 2\alpha_1 -
2\cosh 2 \alpha_2 - \cosh 2 \beta_1 + 2\cosh 2 \beta_2) \nonumber \\
P_4 &=& \frac{2\pi m}{\kappa^2} l_1 l_2 l_3 l_4 V_3 (\cosh 2\alpha_1 -
2\cosh 2 \alpha_2 + 2\cosh 2 \beta_1 - \cosh 2 \beta_2) \nonumber \\
P_5 &=& \frac{2\pi m}{\kappa^2} l_1 l_2 l_3 l_4 V_3 (\cosh 2\alpha_1 +
\cosh 2 \alpha_2 - \cosh 2 \beta_1 - \cosh 2 \beta_2) 
\label{7}
\end{eqnarray}
Eqs. (\ref{5}-\ref{7}) suggest that the gauge charges, ADM mass and the pressures $P_i, 
i=1,...,5$, can replace the ten parameters in the solution (\ref{1}):
$m, \alpha_i, \beta_i, i = 1,2, l_j, j=1,...,4,$ and $V_3$. However, the
pressures $P_i$ are restricted by the following two constraints:
\begin{equation}
P_1 + P_4 + P_5 = 0, \quad P_2 + P_3 + P_5 = 0
\label{8}
\end{equation}
which indicate the exchange symmetry with respect to the two 2-branes
$(\sigma_1 \rightleftharpoons \sigma_2)$, and the two 5-branes $(\omega_1
\rightleftharpoons \omega_2)$, so that the number of independent pressures is
5-2=3. In fact, thanks to these two constraints, three unconstrained 
pressures, the ADM mass and the four gauge charges can match the eight ``numbers'' of
two 2-branes (anti-2-branes) and two 5-branes (anti-5-branes). To see how this
matching works, we calculate the values of the ADM mass and pressures for 
each type of brane in the four extremal limits: $m \to 0, \alpha_i(\beta_i)
\to \pm \infty$ with $q_e^{(i)}(q_m^{(i)})$ and $\alpha_j(\beta_j) (j \not=
i)$ fixed. For one first 2-brane $(\alpha_1 \to \pm \infty)$, the ADM mass and 
pressures reduce to
\begin{equation}
M_{ADM} = - \frac12 P_1 = - \frac12 P_2 = P_3
= P_4 = P_5 = \frac{l_1l_2\hat{e}}{\sqrt{2}\kappa}
\label{9}
\end{equation}
while for the second 2-brane $(\alpha_2 \to \pm \infty)$, they become
\begin{equation}
M_{ADM} = P_1 = P_2 = - \frac12 P_3 = - \frac12 P_4
= P_5 = \frac{l_3l_4\hat{e}}{\sqrt{2}\kappa}
\label{10}
\end{equation}
For a single first 5-brane $(\beta_1 \to \pm \infty)$, one obtains
\begin{equation}
M_{ADM} = -P_1 = \frac12 P_2 = - P_3 = \frac12 P_4
= -P_5 = \frac{l_1l_3V_3}{\sqrt{2}\kappa} \hat{e}_m
\label{11}
\end{equation}
and for the single second 5-brane $(\beta_2 \to \pm \infty)$ they can 
be written as
\begin{equation}
M_{ADM} = \frac12 P_1 = - P_2 = \frac12 P_3 = - P_4
= -P_5 = \frac{l_2l_4V_3}{\sqrt{2}\kappa} \hat{e}_m
\label{12}
\end{equation}
where $\hat{e}, \hat{e}_m$ are the unit charges of the 2-brane and 
the 5-brane \cite{24}
\begin{eqnarray}
\hat{e} &=& \sqrt{2} (2\kappa\pi^2)^{1/3} , \nonumber \\
\hat{e}_m &=& \sqrt{2} (\pi/2\kappa)^{1/3}
\label{13}
\end{eqnarray}
which satisfy the Dirac condition:
\begin{equation}
\hat{e} \cdot \hat{e}_m = 2\pi
\label{14}
\end{equation}
Comparing Eqs. (\ref{5}-\ref{8}) with (\ref{9}-\ref{12}), we can trade eight
parameters of the original ten parameters in the solution (\ref{1}) for
the eight ``numbers'': $N_2^{(1)}, \bar{N}_2^{(1)}, N_2^{(2)}, \bar{N}_2^{(2)},
N_5^{(1)},\linebreak  \bar{N}_5^{(1)},  N_5^{(2)}, \bar{N}_5^{(2)}$, which are the
``numbers'' of 2-branes, anti-2-branes, 5-branes, and anti-5-branes. 
In particular, this can be done by matching three unconstrained pressures, the
ADM mass and four gauge charges to those of a collection of noninteracting
branes. The definitions of the $N$'s are
\begin{eqnarray}
N_2^{(1)} &=& \frac{\sqrt{2}\pi m l_3l_4V_3}{\kappa\hat{e}} e^{2\alpha_1},\quad
\bar{N}_2^{(1)} = \frac{\sqrt{2}\pi m l_3l_4V_3}{\kappa\hat{e}} e^{-2\alpha_1},
\nonumber \\
N_2^{(2)} &=& \frac{\sqrt{2}\pi m l_1l_2V_3}{\kappa\hat{e}} e^{2\alpha_2},\quad
\bar{N}_2^{(2)} = \frac{\sqrt{2}\pi m l_1l_2V_3}{\kappa\hat{e}} e^{-2\alpha_2},
\nonumber \\
N_5^{(1)} &=& \frac{\sqrt{2}\pi m l_2l_4}{\kappa\hat{e}_m} e^{2\beta_1},\quad
\bar{N}_5^{(1)} = \frac{\sqrt{2}\pi m l_2l_4}{\kappa\hat{e}_m} e^{-2\beta_1},
\nonumber \\
N_5^{(2)} &=& \frac{\sqrt{2}\pi m l_1l_3}{\kappa\hat{e}_m} e^{2\beta_2},\quad
\bar{N}_5^{(2)} = \frac{\sqrt{2}\pi m l_1l_3}{\kappa\hat{e}_m} e^{-2\beta_2}
\label{15}
\end{eqnarray}
In terms of these ``numbers'' of M-branes and anti-M-branes, the gauge charges are
\begin{eqnarray}
q_e^{(i)} &=& (N_2^{(i)} - \bar{N}_2^{(i)}) \hat{e} \nonumber \\
q_m^{(i)} &=& (N_5^{(i)} - \bar{N}_5^{(i)}) \hat{e}_m, \quad i=1,2,
\label{16}
\end{eqnarray}
the ADM mass is
\begin{eqnarray}
M_{ADM} = \frac{l_1l_2\hat{e}}{\sqrt{2}\kappa} \left(N_2^{(1)} +
\bar{N}_2^{(1)}\right) + \frac{l_3l_4\hat{e}}{\sqrt{2}\kappa}
\left(N_2^{(2)} +\bar{N}_2^{(2)}\right) \nonumber \\
+ \frac{l_1l_3V_3\hat{e}_m}{\sqrt{2}\kappa} \left(N_5^{(1)} +
\bar{N}_5^{(1)}\right) + \frac{l_2l_4V_3\hat{e}_m}{\sqrt{2}\kappa}
\left(N_5^{(2)} +\bar{N}_5^{(2)}\right),
\label{17}
\end{eqnarray}
and the other parameters are
\begin{equation}
V_3 = \frac{\hat{e}}{e_m} \left(\frac{N_2^{(1)}\bar{N}_2^{(1)} 
N_2^{(2)}\bar{N}_2^{(2)}}{N_5^{(1)}\bar{N}_5^{(1)} N_5^{(2)}\bar{N}_5^{(2)}}
\right)^{\frac14}
\label{18}
\end{equation}
\begin{equation}
l_3 = \left(\frac{N_2^{(1)}\bar{N}_2^{(1)} 
N_5^{(2)}\bar{N}_5^{(2)}}{N_2^{(2)}\bar{N}_2^{(2)} N_5^{(1)}\bar{N}_5^{(1)}}
\right)^{\frac14} l_2
\label{19}
\end{equation}
\begin{equation}
l_4 = \left(\frac{N_2^{(1)}\bar{N}_2^{(1)} 
N_5^{(1)}\bar{N}_5^{(1)}}{N_2^{(2)}\bar{N}_2^{(2)} N_5^{(2)}\bar{N}_5^{(2)}}
\right)^{\frac14} l_1,
\label{20}
\end{equation}
\begin{equation}
m = \frac{\kappa\hat{e}_m}{\sqrt{2}\pi} \left(\frac{N_2^{(2)}\bar{N}_2^{(2)} 
N_5^{(1)}\bar{N}_5^{(1)}N_5^{(2)}\bar{N}_5^{(2)}}{N_2^{(1)}\bar{N}_2^{(1)}}
\right)^{\frac14} (l_1l_2)^{-1}
\label{21}
\end{equation}
Eqs.(\ref{15}-\ref{21}) show that the original ten parameters in the
solution (\ref{1}) can be replaced by eight ``numbers'' of two 2-branes and
anti-2-branes, two 5-branes and anti-5-branes, plus the moduli $l_1, l_2$.
In particular, from Eqs. (\ref{19}, \ref{20}) we find that there are two
constraints among the values of moduli $l_i, i=1,...,4$. Different from the
p-brane solutions in $D = 10$ string theories \cite{8,9}, here even the number 
of branes of each type is given and the values of moduli $l_i$ are not completely
fixed, which is consistent with the black hole picture and resolves the 
discrepancy in the number of free parameters in the D-brane picture. Since the four
M-branes are orthogonal and the mass of each M-brane is proportional to the
common mass parameter $m$, such nonextremal intersecting M-brane configurations
can be viewed as ``bound-state'' configurations.

Upon toroidal compactification to four dimensions, the solution (\ref{1}) is 
reduced to the following Einstein-frame metric \cite{7}
\begin{equation}
ds^2_4 = - f(R) \left(1 - \frac{2m}{R}\right) dt^2 + f^{-1}(R) \left[\left(1-
\frac{2m}{R}\right)^{-1} dR^2+R^2d\Omega^2_2\right]
\label{22}
\end{equation}
with
\begin{equation}
f(R) = \frac{R^2}{\left[(R+2m \sinh^2\alpha_1)(R+2m \sinh^2\alpha_2) 
(R+2m \sinh^2\beta_1) (R+2m \sinh^2\beta_2)\right]^{\frac12}}
\label{23}
\end{equation}
From (\ref{1}), (\ref{22}) and (\ref{23}), we get the Bekenstein-Hawking
entropy
\begin{equation}
S_{BH} = \frac{2\pi A_9}{\kappa^2} = \frac{2\pi A_2}{\kappa^2_4} =
\frac{32\pi^2m^2}{\kappa^2} l_1l_2l_3l_4V_3 \cosh\alpha_1 \cosh\alpha_2
\cosh\beta_1 \cosh\beta_2
\label{24}
\end{equation}
where A is the area of the horizon $(R=0)$, and $\kappa^2_4/8\pi$ is 
Newton's constant in $D=4$ dimensions. The Hawking temperature can be read
off from Eqs. (\ref{22}, \ref{23})
\begin{equation}
T_H = (8\pi m \cosh\alpha_1 \cosh\alpha_2 \cosh\beta_1 \cosh \beta_2)
^{-1}
\label{25}
\end{equation}
In terms of N's, the black hole entropy (\ref{24}) is reexpressed as
\begin{equation}
S_{BH} = 2\pi \left(\sqrt{N_2^{(1)}}+\sqrt{\bar{N}_2^{(1)}}\right)
\left(\sqrt{N_2^{(2)}}+\sqrt{\bar{N}_2^{(2)}}\right)
\left(\sqrt{N_5^{(1)}}+\sqrt{\bar{N}_5^{(1)}}\right)
\left(\sqrt{N_5^{(2)}}+\sqrt{\bar{N}_5^{(2)}}\right)
\label{26}
\end{equation}
which shows that there are ten parameters in the solution (\ref{1}) which
can be replaced by $N_2^{(i)}, \bar{N}_2^{(i)},N_5^{(i)}, \bar{N}_5^{(i)},
 i = 1,2$ plus $l_1, l_2$, the black hole entropy being independent of the
moduli $l_1, l_2$, and this topological character of the entropy is consistent with
that emphasized in \cite{25,26}.

\subsection{The ``boosted'' $5\perp5\perp5$ configuration}
The second nonextremal intersecting M-brane solution in $D=11$ dimensions
corresponds to three 5-branes, each pair intersecting at a 3-brane, with a
boost along a string common to three 3-branes. The nonextremal background
for the $5\perp5\perp5$ configuration with a boost is given by
\begin{eqnarray}
ds^2_{11} &=& (\omega_1\omega_2\omega_3)^{\frac23} \bigg\{(\omega_1\omega_2\omega_3)
^{-1} \left[(-\eta^{-1} \left(1-\frac{2m}{R}\right) dt^2 + \eta (dz_1^{\prime})^2
\right] \nonumber \\
& &+ (\omega_2\omega_3)^{-1}(dz^2_2 + dz^2_3) + (\omega_3\omega_1)^{-1}
(dz^2_4 + dz^2_5) + (\omega_1\omega_2)^{-1}(dz^2_6 + dz^2_7) \nonumber \\ 
& &+\left(1-\frac{2m}{R}\right)^{-1} dR^2 + R^2d\Omega^2_2 \bigg\}, \nonumber \\
F_4 &=& F_4(\omega_1) + F_4(\omega_2) + F_4(\omega_3)
\label{27}
\end{eqnarray}
with
\begin{eqnarray}
F_4(\omega_1) &=& 3(*d\tilde{\omega}_1) \wedge dz_2 \wedge dz_3,
\nonumber \\
F_4(\omega_2) &=& 3(*d\tilde{\omega}_2) \wedge dz_4 \wedge dz_5,
\nonumber \\
F_4(\omega_3) &=& 3(*d\tilde{\omega}_3) \wedge dz_6 \wedge dz_7,
\label{28}
\end{eqnarray}
where the harmonic functions $\eta, \omega_i (\tilde{\omega}_i)$ and the 
differential $dz_1'$ are defined as
\begin{eqnarray}
\eta &=& 1 + \frac{2m \sinh^2\alpha}{R}, \nonumber \\
\omega_i &=& 1 + \frac{2m \sinh^2\beta_i}{R}, \quad \tilde{\omega}_i
= 1 + \frac{m \sinh 2\beta_i}{R}, \quad i = 1,2,3 \nonumber \\
dz'_1 &=& dz_1 - \frac{m \sinh2\alpha}{R} \eta dt
\label{29}
\end{eqnarray}
The electric and three magnetic charges are \cite{27}
\begin{eqnarray}
Q &=& m \sinh 2 \alpha = \frac{2\kappa^2_4}{\omega_2} \cdot
\frac{2\pi(n-\bar{n})}{l_1} = \frac{\kappa^2(n-\bar{n})}{l^2_1S_{23}
S_{45}S_{67}} \nonumber \\
q_m^{(1)} &=& \frac{1}{\sqrt{2}\kappa} \int F_4(\omega_1) =
\frac{4\pi m \sinh2\beta_1}{\sqrt{2}\kappa} S_{23} \nonumber \\
q_m^{(2)} &=& \frac{1}{\sqrt{2}\kappa} \int F_4(\omega_2) =
\frac{4\pi m \sinh2\beta_2}{\sqrt{2}\kappa} S_{45} \nonumber \\
q_m^{(3)} &=& \frac{1}{\sqrt{2}\kappa} \int F_4(\omega_3) =
\frac{4\pi m \sinh2\beta_3}{\sqrt{2}\kappa} S_{67} 
\label{30}
\end{eqnarray}
where $n, \bar{n}$ are positive integers and the brane-coordinates
$z_1, z_{2,3}, z_{4,5}, z_{6,7}$ are compactified on $S^1 \times T^2\times T^2
\times T^2$, the circumference of circle and areas of three 2-tori being
$l_1, S_{23}, S_{45}, S_{67}$ respectively, and the solution (\ref{27}) has
nine parameters: $m, \alpha, \beta_i, i = 1,2,3, l_1, S_{23}, S_{45},
S_{67}$.

The ADM mass and the pressures of the solution (\ref{27}) are found to be
\begin{eqnarray}
M_{ADM} &=& \frac{2\pi m}{\kappa^2}l_1S_{23}S_{45}S_{67}(\cosh 2\alpha +
\cosh 2 \beta_1 + \cosh 2\beta_2 + \cosh2\beta_3), \nonumber \\
P_1 &=& \frac{2\pi m}{\kappa^2}l_1S_{23}S_{45}S_{67}(3\cosh 2\alpha -
\cosh 2 \beta_1 - \cosh 2\beta_2 - \cosh2\beta_3), \nonumber \\
P_{23} &=& \frac{2\pi m}{\kappa^2}l_1S_{23}S_{45}S_{67}(2
\cosh 2 \beta_1 - \cosh 2\beta_2 - \cosh2\beta_3), \nonumber \\
P_{45} &=& \frac{2\pi m}{\kappa^2}l_1S_{23}S_{45}S_{67}(-
\cosh 2 \beta_1 + 2\cosh 2\beta_2 - \cosh2\beta_3), \nonumber \\
P_{67} &=& \frac{2\pi m}{\kappa^2}l_1S_{23}S_{45}S_{67}(-
\cosh 2 \beta_1 - \cosh 2\beta_2 + 2\cosh2\beta_3),
\label{31} 
\end{eqnarray}
Eqs. (\ref{30}, \ref{31}) suggest that the nine parameters in the 
solution (\ref{27}) can be replaced by other nine quantities:
$Q, q_m^{(i)}, i = 1,2,3, P_1, P_{23}, P_{45}, P_{67}$. But the pressures
are not independent and are related through the constraint
\begin{equation}
P_{23} + P_{45} + P_{67} = 0
\label{32}
\end{equation}
which reflects the exchange symmetry of the solution (\ref{27}) with 
respect to three 5-branes, so that the number of unconstrained pressures is
$4-1=3$. Following a discussion similar to that of section 2.1, the ``numbers'' of
three 5-branes, anti-5-branes, right-moving strings and left-moving strings
are defined by
\begin{eqnarray}
N_5^{(1)} &=& \frac{\sqrt{2}\pi m S_{23}}{\kappa \hat{e}_m} e^{2\beta_1},
\quad \bar{N}_5^{(1)} = \frac{\sqrt{2}\pi m S_{23}}{\kappa \hat{e}_m} e^{-2\beta_1},
\nonumber \\
N_5^{(2)} &=& \frac{\sqrt{2}\pi m S_{45}}{\kappa \hat{e}_m} e^{2\beta_2},
\quad \bar{N}_5^{(2)} = \frac{\sqrt{2}\pi m S_{45}}{\kappa \hat{e}_m} e^{-2\beta_2},
\nonumber \\
N_5^{(3)} &=& \frac{\sqrt{2}\pi m S_{67}}{\kappa \hat{e}_m} e^{2\beta_3},
\quad \bar{N}_5^{(3)} = \frac{\sqrt{2}\pi m S_{67}}{\kappa \hat{e}_m} e^{-2\beta_3},
\nonumber \\
n &=& \frac{ml^2_1S_{23}S_{45}S_{67}}{2\kappa^2} e^{2\alpha}, \quad
\bar{n} = \frac{ml^2_1S_{23}S_{45}S_{67}}{2\kappa^2} e^{-2\alpha}
\label{33}
\end{eqnarray}
The gauge charges are
\begin{eqnarray}
q_m^{(i)} &=& \left(N_5^{(i)} - \bar{N}_5^{(i)}\right) \hat{e}_m, \quad
i = 1,2,3
\nonumber \\
q_e &=& (n-\bar{n}) \hat{e},
\label {34}
\end{eqnarray}
and the ADM mass is
\begin{eqnarray}
M_{ADM} &=& \frac{\hat{e}_m}{\sqrt{2}\kappa} \bigg[ l_1S_{45}S_{67}
\left(N_5^{(1)} + \bar{N}_5^{(1)}\right)+ l_1S_{23}S_{67}
\left(N_5^{(2)} + \bar{N}_5^{(2)}\right) \nonumber \\
& & l_1S_{23}S_{45} \left(N_5^{(3)} + \bar{N}_5^{(3)}\right) + \frac
{2\sqrt{2}\pi\kappa}{\hat{e}_ml_1} (n+\bar{n}) \bigg],
\label{35}
\end{eqnarray}
and the values of the moduli are
\begin{eqnarray}
l^2_1 &=& \frac{\sqrt{2}\pi m^2}{\kappa \hat{e}^3_m} \left(\frac{n\bar{n}}
{\prod\limits^3_{i=1} N_5^{(i)} \bar{N}_5^{(i)}} \right)^{\frac12}, \nonumber \\
S_{23} &=& \frac{\kappa\hat{e}_m}{\sqrt{2}\pi} \sqrt{N_5^{(1)} \bar{N}_5^{(1)}}
\,m^{-1}, \nonumber \\
S_{45} &=& \frac{\kappa\hat{e}_m}{\sqrt{2}\pi} \sqrt{N_5^{(2)} \bar{N}_5^{(2)}}
\,m^{-1}, \nonumber \\
S_{67} &=& \frac{\kappa\hat{e}_m}{\sqrt{2}\pi} \sqrt{N_5^{(3)} \bar{N}_5^{(3)}}
\,m^{-1}, 
\label{36}
\end{eqnarray}
which show that the moduli depend not only on the ``numbers'' of M-branes
and anti-M-branes, but also on the nonextremality parameter $m$. Therefore, when the
``numbers'' of M-branes and anti-M-branes are kept fixed, the values of the moduli cannot
be determined completely, which reconciles the apparent contradiction between the number of free
parameters of the D-brane picture and those of the black hole one. Furthermore,
from (\ref{33}-\ref{36}) we find that the eight ``numbers'' of the three 5-branes,
anti-5-branes, right-moving strings and left-moving strings, plus the nonextremality
parameter $m$ can replace the original nine parameters in solution
(\ref{27}).

When compactified to four dimensions, the solution (\ref{27}) is reduced to 
(\ref{22}) with
\begin{equation}
f(R) = R^2\left[(R+2m \sinh^2\alpha) (R+2m \sinh^2\beta_1)(R+2m \sinh^2\beta_2)
(R+2m \sinh^2\beta_3)\right]^{-\frac12}
\label{37}
\end{equation}
Then the Bekenstein-Hawking entropy can derived from (\ref{22}) and
(\ref{37}) to be
\begin{equation}
S_{BH} = \frac{2\pi A_9}{\kappa^2} = \frac{2\pi A_2}{\kappa^2_4} =
\frac{32\pi^2m^2}{\kappa^2} l_1S_{23}S_{45}S_{67} \cosh\alpha  
\cosh\beta_1 \cosh \beta_2 \cosh \beta_3
\label{38}
\end{equation}
and the Hawking temperature is
\begin{equation}
T_H = (8\pi m \cosh\alpha \cosh\beta_1 \cosh\beta_2 \cosh\beta_3)^{-1}
\label{39}
\end{equation}
In terms of N's, Eq. (\ref{38}) can be rewritten as 
\begin{equation}
S_{BH} = 2\pi \left( \sqrt{N_5^{(1)}} + \sqrt{\bar{N}_5^{(1)}}\right)
\left( \sqrt{N_5^{(2)}} + \sqrt{\bar{N}_5^{(2)}}\right)
\left( \sqrt{N_5^{(3)}} + \sqrt{\bar{N}_5^{(3)}}\right)
\left( \sqrt{n} + \sqrt{\bar{n}}\right)
\label{40}
\end{equation}
which shows that even here there are nine free parameters: $n, \bar{n}, N_5^{(i)},
\bar{N}_5^{(i)}, i=1,2,3$, plus $m$, the Bekenstein-Hawking entropy being 
independent of the parameter $m$. The same prefactors in (\ref{26})
and (\ref{40}) imply that the above two nonextremal $4D$ black hole
backgrounds obtained from $2\perp2\perp5\perp5$ and ``boosted'' $5\perp5\perp5$ 
nonextremal intersecting M-brane configurations can be related by the 
symmetry transformation of M-theory, which is a combination of T-duality
and SL(2,Z) symmetry of $D = 10$ type IIB theory lifted to $D = 11$ M-theory.

\section{Nonextremal 5D black holes}
The nonextremal $5D$ black holes with three independent charges can be
obtained from two different nonextremal intersecting M-brane solutions:
$2\perp2\perp2$ (three 2-branes intersecting at a point), and ``boosted''
$2\perp5$ (intersecting 2-brane and 5-brane with a momentum along the
common string).

\subsection{The $2\perp2\perp2$, $D = 11$ solution}
The nonextremal intersecting M-brane solution corresponding to the 
$2\perp2\perp2$ configuration is given by
\begin{eqnarray*}
ds^2_{11} &=& (\sigma_1\sigma_2\sigma_3)^{\frac13} \bigg[-(\sigma_1
\sigma_2\sigma_3)^{-1} \left(1 - \frac{2m}{R^2}\right) dt^2 + \sigma_1^{-1}
(dz^2_1+dz^2_2) \nonumber \\
& &+ \sigma_2^{-1} (dz^2_3+dz^2_4) + \sigma_3^{-1}(dz^2_5+dz^2_6) + 
\left(1 - \frac{2m}{R^2}\right)^{-1} dR^2+R^2d\Omega^2_3 \bigg] 
\end{eqnarray*}
and
\begin{equation}
F_4 = F_4(\sigma_1) + F_4(\sigma_2) + F_4(\sigma_3)
\label{41}
\end{equation}
with
\begin{eqnarray}
F_4(\sigma_1) &=& 3dt \wedge (\sigma_1^{-2} d\tilde{\sigma}_1) \wedge
dz_1 \wedge dz_2, \nonumber \\
F_4(\sigma_2) &=& 3dt \wedge (\sigma_2^{-2} d\tilde{\sigma}_2) \wedge
dz_3 \wedge dz_4, \nonumber \\
F_4(\sigma_3) &=& 3dt \wedge (\sigma_3^{-2} d\tilde{\sigma}_3) \wedge
dz_5 \wedge dz_6,
\label{42}
\end{eqnarray}
where the harmonic functions $\sigma_i, \tilde{\sigma}_i$ for the three
2-branes are 
\begin{eqnarray}
\sigma_i &=& 1 + \frac{2m \sinh^2\alpha_i}{R^2}, \nonumber \\
\tilde{\sigma}_i &=& 1 + \frac{m \sinh 2\alpha_i}{R^2}, \quad i=1,2,3
\label{43}
\end{eqnarray}
Eqs. (\ref{42}, \ref{43}) show that the expressions for the field strengths 
$F_4(\sigma_i)$ are different from those in \cite{7}; here two sets of 
harmonic functions, $\sigma_i$ and $\tilde{\sigma}_i$, are exploited to 
describe $F_4(\sigma_i)$. The dual forms in 11 dimensions of $F_4(\sigma_i)$
are given by
\begin{eqnarray}
^{\hat{*}}F_4(\sigma_1) &=& 2m (\sinh2\alpha_1) \epsilon_3 \wedge dz_3 
\wedge dz_4 \wedge dz_5 \wedge dz_6, \nonumber \\
^{\hat{*}}F_4(\sigma_2) &=& 2m (\sinh2\alpha_2) \epsilon_3 \wedge dz_1 
\wedge dz_2 \wedge dz_5 \wedge dz_6, \nonumber \\
^{\hat{*}}F_4(\sigma_3) &=& 2m (\sinh2\alpha_3) \epsilon_3 \wedge dz_1 
\wedge dz_2 \wedge dz_3 \wedge dz_4, 
\label{44}
\end{eqnarray}
where $\epsilon_3$ is the volume 3-form of $S^3$. The electric charges
are then
\begin{eqnarray}
q_e^{(1)} = \frac{1}{\sqrt{2}\kappa} \int\, ^{\hat{*}}F_4(\sigma_1) =
\frac{4\pi^2m \sinh2\alpha_1}{\sqrt{2}\kappa} S_{34}S_{56}, \nonumber \\
q_e^{(2)} = \frac{1}{\sqrt{2}\kappa} \int\, ^{\hat{*}}F_4(\sigma_2) =
\frac{4\pi^2m \sinh2\alpha_2}{\sqrt{2}\kappa} S_{12}S_{56}, \nonumber \\
q_e^{(3)} = \frac{1}{\sqrt{2}\kappa} \int\, ^{\hat{*}}F_4(\sigma_3) =
\frac{4\pi^2m \sinh2\alpha_3}{\sqrt{2}\kappa} S_{12}S_{34}, 
\label{45}
\end{eqnarray}
where the internal coordinates $z_{1,2}, z_{3,4}, z_{5,6}$ are compactified
on $T^2 \times T^2\times T^2$, and the areas of three 2-tori are
$S_{12}, S_{34}, S_{56}$ respectively. Thus the solution (\ref{41})
contains seven parameters: $m, \alpha_i, i = 1,2,3, S_{12}, S_{34}, S_{56}$.

The ADM mass and the pressures of the solution (\ref{41}) are
\begin{eqnarray}
M_{ADM} &=& \frac{2\pi^2m}{\kappa_2} S_{12}S_{34}S_{56} (\cosh2\alpha_1 +
\cosh2\alpha_2 + \cosh2\alpha_3) \nonumber \\
P_{12} &=& \frac{2\pi^2m}{\kappa_2} S_{12}S_{34}S_{56} (-2\cosh2\alpha_1 +
\cosh2\alpha_2 + \cosh2\alpha_3) \nonumber \\
P_{34} &=& \frac{2\pi^2m}{\kappa_2} S_{12}S_{34}S_{56} (\cosh2\alpha_1 
-2\cosh2\alpha_2 + \cosh2\alpha_3) \nonumber \\
P_{56} &=& \frac{2\pi^2m}{\kappa_2} S_{12}S_{34}S_{56} (\cosh2\alpha_1 +
\cosh2\alpha_2 - 2\cosh2\alpha_3) 
\label{46}
\end{eqnarray}

As discussed in former sections, the pressures are related to each other by 
the constraint
\begin{equation}
P_{12} + P_{34} + P_{56} = 0
\label{47}
\end{equation}
which shows the exchange symmetry of the solution (\ref{41}) with respect to 
the three 2-branes. Due to the presence of the constraint (\ref{47}), the
number of unconstrained pressures is 2. Then two unconstrained pressures,
the ADM mass and the three electric charges match the six ``numbers'' of three 2-branes
and 2-antibranes. Following the similar procedure of section 2.1, but only
taking three extremal limits: $m \to 0, \alpha_i \to \pm \infty$ with
$q_e^{(i)}$ and $\alpha_j (j \neq i)$ fixed, we can define the ``numbers''
of three 2-branes and anti-2-branes as follows
\begin{eqnarray}
N_2^{(1)} &=& \frac{\sqrt{2}\pi^2m S_{34}S_{56}}{\kappa \hat{e}} e^{2\alpha_1},
\quad \bar{N}_2^{(1)} = \frac{\sqrt{2}\pi^2m S_{34}S_{56}}{\kappa \hat{e}} 
e^{-2\alpha_1}, \nonumber \\
N_2^{(2)} &=& \frac{\sqrt{2}\pi^2m S_{12}S_{56}}{\kappa \hat{e}} e^{2\alpha_2},
\quad \bar{N}_2^{(2)} = \frac{\sqrt{2}\pi^2m S_{12}S_{56}}{\kappa \hat{e}} 
e^{-2\alpha_2}, \nonumber \\
N_2^{(3)} &=& \frac{\sqrt{2}\pi^2m S_{12}S_{34}}{\kappa \hat{e}} e^{2\alpha_3},
\quad \bar{N}_2^{(3)} = \frac{\sqrt{2}\pi^2m S_{12}S_{34}}{\kappa \hat{e}} 
e^{-2\alpha_3}. 
\label{48}
\end{eqnarray}
Then the three electric charges are simply
\begin{equation}
q_e^{(i)} = \left(N_2^{(i)} - \bar{N}_2^{(i)}\right) \hat{e}, \; i = 1,2,3
\label{49}
\end{equation}
and the ADM mass is
\begin{equation}
M_{ADM} = \frac{\hat{e}}{\sqrt{2}\kappa} \left[S_{12} \left(N_2^{(1)} + \bar{N}_2^{(1)}\right)
+ S_{34} \left(N_2^{(2)} + \bar{N}_2^{(2)}\right) + S_{56}
\left(N_2^{(3)} + \bar{N}_2^{(3)}\right) \right]
\label{50}
\end{equation}
and the values of moduli are
\begin{eqnarray}
S_{12} &=& \left(\frac{\kappa\hat{e}}{\sqrt{2}\pi^2}\right)^{\frac12}
\left( \frac{N_2^{(2)}\bar{N}_2^{(2)}N_2^{(3)}\bar{N}_2^{(3)}}
{N_2^{(1)}\bar{N}_2^{(1)}} \right)^{\frac14} m^{-\frac12}, \nonumber \\
S_{34} &=& \left(\frac{\kappa\hat{e}}{\sqrt{2}\pi^2}\right)^{\frac12}
\left( \frac{N_2^{(3)}\bar{N}_2^{(3)}N_2^{(1)}\bar{N}_2^{(1)}}
{N_2^{(2)}\bar{N}_2^{(2)}} \right)^{\frac14} m^{-\frac12}, \nonumber \\
S_{56} &=& \left(\frac{\kappa\hat{e}}{\sqrt{2}\pi^2}\right)^{\frac12}
\left( \frac{N_2^{(1)}\bar{N}_2^{(1)}N_2^{(2)}\bar{N}_2^{(2)}}
{N_2^{(3)}\bar{N}_2^{(3)}} \right)^{\frac14} m^{-\frac12}, 
\label{51}
\end{eqnarray}
which means that the values of the moduli are functions of the nonextremality
parameter m and the ``numbers'' of 2-branes and anti-2-branes, which is
different from the situation in the D-brane picture. Eqs. (\ref{48} -
\ref{51}) reveal that the original seven parameters in the solution
(\ref{41}) can be replaced by six ``numbers'' of three 2-branes and anti-2-branes
plus the parameter $m$.

The five-dimensional Einstein-frame metric obtained by reduction of the
internal coordinates $z_1,...,z_6$ is
\begin{equation}
ds^2_5 = - f^2(R)\left(1 - \frac{2m}{R^2}\right) dt^2 + f^{-1}(R)
\left[\left(1 - \frac{2m}{R^2}\right)^{-1} dR^2 + R^2d\Omega^2_3 \right]
\label{52}
\end{equation}
with
\begin{equation}
f(R) = \frac{R^2}{\left[(R^2+2m \sinh^2\alpha_1) (R^2+2m \sinh^2\alpha_2)
(R^2+2m \sinh^2\alpha_3)\right]^{\frac13}}
\label{53}
\end{equation}
Now the Bekenstein-Hawking entropy can be obtained from (\ref{41}),
(\ref{52}) and (\ref{53}) and is found to be 
\begin{equation}
S_{BH} = \frac{2\pi A_9}{\kappa^2} = \frac{2\pi A_3}{\kappa^2_5}
= \frac{8\sqrt{2}\pi^3m^{3/2}}{\kappa^2} S_{12}S_{34}S_{56} \times
\cosh\alpha_1 \times \cosh \alpha_2 \times \cosh \alpha_3
\label{54}
\end{equation}
where $\kappa^2_5/8\pi$ is Newton's constant in $D=5$ dimensions, and the
Hawking temperature is
\begin{equation}
T_H = \left(2\pi\sqrt{2m} \cosh \alpha_1 \cosh\alpha_2 \cosh\alpha_3\right)^{-1}
\label{55}
\end{equation}
In terms of $N$'s, Eq. (\ref{54}) can be rewritten as
\begin{equation}
S_{BH} = 2\pi \left(\sqrt{N_2^{(1)}} + \sqrt{\bar{N}_2^{(1)}}\right)
\left(\sqrt{N_2^{(2)}} + \sqrt{\bar{N}_2^{(2)}}\right)
\left(\sqrt{N_2^{(3)}} + \sqrt{\bar{N}_2^{(3)}}\right)
\label{56}
\end{equation}
which shows that the black hole entropy depends only on the ``numbers'' of
2-branes and anti-2-branes, but is independent of the nonextremality parameter
$m$.

\subsection{The ``boosted'' $2\perp5$ configuration}
The configuration of a 2-brane intersecting with a 5-brane with a ``boost'' along the
common string is given by
\begin{eqnarray*}
ds^2_{11} &=& \sigma^{\frac13} \omega^{\frac23} \bigg\{ (\sigma\omega)^{-1}
\left[-\eta^{-1} \left(1-\frac{2m}{R^2}\right) dt^2+\eta dz_1'^2\right]
\nonumber \\
& & \sigma^{-1} dz^2_2 + \omega^{-1} (dz^2_3 + dz^2_4 + dz^2_5 + dz^2_6)
\nonumber \\
& &+ \left(1-\frac{2m}{R^2}\right)^{-1} dR^2 + R^2d\Omega^2_3 \bigg\}
\end{eqnarray*}
with
\begin{equation}
F_4 = F_4(\sigma) + F_4(\omega)
\label{57}
\end{equation}
where
\begin{eqnarray}
F_4(\sigma) &=& 3dt \wedge (\sigma^{-2}d\tilde{\sigma}) \wedge dz'_1
\wedge dz_2, \nonumber \\
F_4(\omega) &=& 3(*d\tilde{\omega}) \wedge dz_2
\label{58}
\end{eqnarray}
where the harmonic functions $\sigma(\tilde{\sigma}), \omega(\tilde
{\omega}), \eta$ and the differential $dz'_1$ are defined as
\begin{eqnarray}
\sigma &=& 1 + \frac{2m \sinh^2\alpha}{R^2}, \quad \tilde{\sigma}
= 1 + \frac{m \sinh2\alpha}{R^2}, \nonumber \\
\omega &=& 1 + \frac{2m \sinh^2\beta}{R^2}, \quad \tilde{\omega}
= 1 + \frac{m \sinh2\beta}{R^2}, \nonumber \\
\eta &=& 1 + \frac{2m\sinh^2\delta}{R^2}, \nonumber \\
dz'_1 &=& dz_1 - \frac{m \sinh2\delta}{R^2} \eta dt
\label{59}
\end{eqnarray}
Then the electric and magnetic charges are given by
\begin{eqnarray*}
q_e &=& \frac{1}{\sqrt{2}\kappa} \int\,^{\hat{*}}F_4(\sigma) =
\frac{4\pi^2m \sinh2\alpha}{\sqrt{2}\kappa} V_4, \nonumber \\
q_m &=& \frac{1}{\sqrt{2}\kappa} \int F_4(\omega) =
\frac{4\pi^2m \sinh2\beta}{\sqrt{2}\kappa} l_2, 
\end{eqnarray*}
with
\begin{equation}
Q = m \sinh2\delta = \frac{2\kappa^2_5}{2\omega_3} \cdot \frac
{2\pi(n-\bar{n})}{l_1} = \frac{\kappa^2(n-\bar{n})}{\pi l^2_1l_2V_4}
\label{60}
\end{equation}
where $n, \bar{n}$ are positive integers, and the internal coordinates
$z_1, z_2, z_{3,4,5,6}$ are compactified on $S^1\times S^1\times T^4$, the
circumferences of two circles and the volume of the 4-torus being taken to be 
$l_1, l_2, V_4$ respectively. The solution (\ref{57}) therefore has seven
parameters: $m, \alpha, \beta, \delta, l_1, l_2, V_4$.

The corresponding ADM mass and pressures of the solution (\ref{47}) are
found to be
\begin{eqnarray}
M_{ADM} &=& \frac{2\pi^2m}{\kappa^2} l_1l_2V_4 (\cosh2\alpha +
\cosh2\beta + \cosh2\delta), \nonumber \\
P_1 &=& \frac{2\pi^2m}{\kappa^2} l_1l_2V_4 (-2\cosh2\alpha -
\cosh2\beta+ 3\cosh2\delta), \nonumber \\
P_2 &=& \frac{2\pi^2m}{\kappa^2} l_1l_2V_4 (-2\cosh2\alpha +
2\cosh2\beta), \nonumber \\
P_3 &=& \frac{2\pi^2m}{\kappa^2} l_1l_2V_4 (\cosh2\alpha -
\cosh2\beta).
\label{61}
\end{eqnarray}
The above three pressures are not independent, but restricted by the 
constraint
\begin{equation}
P_2 + 2P_3 = 0
\label{62}
\end{equation}
Proceeding along similar lines as in section 2.1, the ``numbers'' of 2-branes 
(anti-2-branes), 5-branes (anti-5-branes) and right (left)-moving strings
can be defined as
\begin{eqnarray}
N_2 &=& \frac{\sqrt{2}\pi^2mV_4}{\kappa\hat{e}} e^{2\alpha}, \quad
\bar{N}_2 = \frac{\sqrt{2}\pi^2mV_4}{\kappa\hat{e}} e^{-2\alpha}, \nonumber \\
N_5 &=& \frac{\sqrt{2}\pi^2ml_2}{\kappa\hat{e}_m} e^{2\beta}, \quad
\bar{N}_5 = \frac{\sqrt{2}\pi^2ml_2}{\kappa\hat{e}_m} e^{-2\beta}, \nonumber \\
n &=& \frac{\pi ml_1^2l_2V_4}{2\kappa^2} e^{2\delta}, \quad
\bar{n} = \frac{\pi ml_1^2l_2V_4}{2\kappa^2} e^{-2\delta}
\label{63}
\end{eqnarray}
In terms of these the electric and magnetic charges are
\begin{equation}
q_e = (N_2-\bar{N}_2)\hat{e}, \; q_m = (N_5-\bar{N}_5)\hat{e}_m,
\; q = (n-\bar{n})\hat{e}.
\label{64}
\end{equation}
Also the ADM mass is
\begin{equation}
M_{ADM} = \frac{1}{\sqrt{2}\kappa} \left[l_1l_2\hat{e}(N_2+\bar{N}_2) +
l_1V_4\hat{e}_m(N_5+\bar{N}_5) + \frac{2\sqrt{2}\kappa\pi}{l_1} (n+\bar{n})
\right] 
\label{65}
\end{equation}
and the values of moduli are
\begin{eqnarray}
l_1 &=& \sqrt{2}\pi \left(\frac{n\bar{n}}{N_2\bar{N}_2N_5\bar{N}_5}
\right)^{\frac14} m^{\frac12}, \nonumber \\
l_2 &=& \frac{\kappa\hat{e}_m}{\sqrt{2}\pi^2} \sqrt{N_5\bar{N}_5} \,m^{-1},
\nonumber \\
V_4 &=& \frac{\kappa\hat{e}}{\sqrt{2}\pi^2} \sqrt{N_2\bar{N}_2} \,m^{-1},
\label{66}
\end{eqnarray}
which displays that there are constraints among the moduli due to the 
dependence of the mass of the 2-brane on that of 5-brane. Eqs. (\ref{63}-\ref{66}) imply that the original seven parameters in (\ref{57}) can be
replaced by six ``numbers'' of 2-branes (anti-2-branes), 5-branes
(anti-5-branes), and right (left)-moving strings, plus the parameter $m$.

Compactifying the brane-coordinates $z_i, i = 1,...,6$, one obtains the
five-dimensional Einstein-frame metric which has the form of (\ref{52}), but
with
\begin{equation}
f(R) = \frac{R^2}{\left[(R^2+2m \sinh^2\alpha) (R^2+2m \sinh^2\beta)
(R^2+2m \sinh^2\delta)\right]^{\frac13}}
\label{67}
\end{equation}
Then the Bekenstein-Hawking entropy is
\begin{equation}
S_{BH} = \frac{2\pi A_9}{\kappa^2} = \frac{2\pi A_3}{\kappa^2_5} =
\frac{8\sqrt{2}\pi^3m^{3/2}}{\kappa^2} l_1l_2V_4 \cdot \cosh\alpha \cdot
\cosh\beta \cdot \cosh\delta,
\label{68}
\end{equation}
and the Hawking temperature is
\begin{equation}
T_H = (2\pi\sqrt{2m} \cosh\alpha \cosh\beta \cosh\delta)^{-1}
\label{69}
\end{equation}
Reexpressing (\ref{68}) in terms of the ``numbers'' of M-branes 
and anti-M-branes, we have
\begin{equation}
S_{BH} = 2\pi \left(\sqrt{N_2}+\sqrt{\bar{N}_2}\right) 
\left(\sqrt{N_5}+\sqrt{\bar{N}_5}\right) \left(\sqrt{n}+\sqrt{\bar{n}}\right)
\label{70}
\end{equation}
which means that the black hole entropy depends only on the ``numbers''
of the 2-(anti)-branes, 5-(anti)-branes, and the right (left)-moving strings,
but is independent of the nonextremality parameter $m$. From (\ref{26}), (\ref{40}),
(\ref{56}) and (\ref{70}), we find that in terms of the ``numbers'' of
M-branes and anti-M-branes, the prefactors of the Bekenstein-Hawking entropies
for nonextremal $4D$ and $5D$ black holes are model independent \cite{28}.

\section{Microscopic explanation}
In M-theory, 5-branes interact via exchange of membranes with boundaries
on the 5-branes \cite{29}-\cite{31}, and quantization of the boundary states results
in the 5-brane low-energy effective theory \cite{18, 32}, which suggests that
collapsed membranes live on the intersection manifold of our intersecting
5-branes and bind them together \cite{22}. As an example, we consider the
``boosted'' $5\perp5\perp5$ M-brane solution, and associate the microscopic
massless states with those of 2-branes attached to 5-branes near the 
intersection point. In particular, one can visualize a 2-brane with three
holes, each of them attached to different 5-dimensional hyperplanes in which the
5-branes lie, and for the three 5-branes intersecting along a string, the 
collapsed membrane gives the desired string, the momentum of the membranes
then becoming the momentum of the string. The generating function of the 
degeneracy of states with momentum $2\pi n/l_1$ is then given by \cite{16,
13}
\begin{equation}
\sum_n d(n)q^n = \frac{\pi_n(1+q^n)^K}{\pi_n(1-q^n)^K}
\label{71}
\end{equation}
with
\begin{equation}
K = 4N_5^{(1)} N_5^{(2)} N_5^{(3)}
\label{72}
\end{equation}
which means that the collapsed membranes act like a single self-dual string
with 4 bosonic and 4 fermionic zero modes. The degeneracy of states is then
$d(n) = \exp \left[2\pi\sqrt{n N_5^{(1)} N_5^{(2)} N_5^{(3)}} \right]$ for
large $n$, and the entropy 
\begin{equation}
S_{stat} = \ln d(n) = 2\pi \sqrt{n N_5^{(1)} N_5^{(2)} N_5^{(3)}}
\label{73}
\end{equation}
Considering now the near-extremal black hole, when $\bar{N}_5^{(1)} =
\bar{N}_5^{(2)} = \bar{N}_5^{(3)} = 0$ and $l_1$ is large, the lightest
excitations will be the momentum modes. At weak coupling, the entropy has the
form
\begin{equation}
S_{stat} = 2\pi \sqrt{N_5^{(1)} N_5^{(2)} N_5^{(3)}} \left(\sqrt{n} +
\sqrt{\bar{n}}\right)
\label{74}
\end{equation}
where $\bar{n}$ denotes the momentum of the left movers. Since U-duality
interchanges the different branes and strings, a result similar to
(\ref{74}) with the indices permuted is obtained for these cases. 
Thus the entropy (\ref{40})
is the simplest duality invariant expression which is compatible with
the different nonextremal limits.

The above algorithm holds for the nonextremal $5D$ black hole obtained upon
toroidal compactification from the ``boosted'' $2\perp5$ nonextremal
intersecting M-brane solution. From the symmetry transformation, which is a
combination of T-duality and SL(2,Z) symmetry of the $D=10$ type IIB string
theory lifted to $D=11$ M-theory, we can obtain the statistical entropy for the 
other two nonextremal black holes obtained from $2\perp2\perp5\perp5$ and 
$2\perp2\perp2$ nonextremal intersecting M-brane configurations.

\section{Conclusion}
In the above, the D-brane and anti-D-brane picture has been generalized to 
an M-brane and anti-M-brane one, and the nonextremal $4D$ and $5D$ black holes 
obtained upon toroidal compactification from the nonextremal intersecting M-brane
solutions have been identified as a collection of M-branes and anti-M-branes.
In our improved nonextremal intersecting M-brane solutions, two sets of
harmonic functions were first introduced, which makes it easy to
explicitly calculate the electric charges. In a unified frame of $D=11$
M-theory, the ``numbers'' of M-branes and anti-M-branes have been defined
consistently. The constraints among the pressures, here found for the first time,
reflect the exchange symmetry of the nonextremal intersecting M-brane
solutions, and only the unconstrained pressures, ADM mass and gauge charges
match the ``numbers'' of M-branes and anti-M-branes. Unlike the D-brane and
anti-D-brane picture in which the moduli depend only on the ``numbers'' of
D-branes and anti-D-branes, the new features in the M-brane and anti-M-brane 
picture have been shown to be: 1) there are constraints among the 
moduli, and some moduli can be chosen as free parameters; 2) the moduli
are functions of the nonextremality parameter $m$ and the ``numbers''
of M-branes and anti-M-branes. As a result, these new features resolve 
the discrepancy in the number of free parameters in the D-brane picture. In
terms of the ``numbers'' of M-branes and anti-M-branes, the prefactors of the
entropies have been found to be model independent, and the Bekenstein-Hawking
entropy assumes the $E_7(E_6)$ invariant form for $4D$ ($5D$) black holes. The
microscopic origin of the Bekenstein-Hawking entropy for the nonextremal
black holes has been explained from the M-brane and anti-M-brane picture. 
All of these features together seem to indicate that the extension of the 
D-brane and anti-D-brane
picture to the M-brane and anti-M-brane one is quite convincing. However, such 
an M-brane and anti-M-brane picture is invalid for the analysis of the
nonextremal black holes in $6 \le D\le9 $ dimensions, where the ``numbers'' of
M-branes and anti-M-branes cannot be defined consistently, and the black hole
entropy depends on the nonextremality parameter $m$.

\section*{Acknowledgements}
This work was supported in part by the European Union Under the Human Capital 
and Mobility Programme. J.--G. Z. thanks the Alexander von Humboldt
Foundation for financial support in the form of a research fellowship.
J.--Q. L. is indebted to the Deutsche Forschungsgemeinschaft for financial
support.

\end{document}